\documentclass[twocolumn,showpacs,preprintnumbers,amsmath,amssymb]{revtex4}

\usepackage{graphicx}% Include figure files
\usepackage{dcolumn}% Align table columns on decimal point
\usepackage{bm}% bold math
\usepackage{amsmath}

\begin{document}
\title{Geographical networks evolving with an optimal policy}
\author{Yan-Bo Xie$^{1}$}
\author{Tao Zhou$^{1,2}$}
\email{zhutou@ustc.edu}
\author{Wen-Jie Bai$^{3}$}
\author{Guanrong Chen$^{2}$}
\author{Wei-Ke Xiao$^{4}$}
\author{Bing-Hong Wang$^{1}$}

\affiliation{%
$^{1}$Department of Modern Physics and Nonlinear Science Center,
University of Science and Technology of China, Hefei 230026,
PR China\\
$^{2}$Department of Electronic Engineering, City University of
Hong Kong, Hong Kong SAR, PR China \\
$^{3}$Department of Chemistry, University of Science and
Technology of China, Hefei 230026, PR China\\
$^{4}$Center for Astrophysics, University of Science and
Technology of China, Hefei 230026, PR China
}%

\date{\today}

\begin{abstract}
In this article, we propose a growing network model based on an
optimal policy involving both topological and geographical
measures. In this model, at each time step, a new node, having
randomly assigned coordinates in a $1 \times 1$ square, is added
and connected to a previously existing node $i$, which minimizes
the quantity $r_i^2/k_i^\alpha$, where $r_i$ is the geographical
distance, $k_i$ the degree, and $\alpha$ a free parameter. The
degree distribution obeys a power-law form when $\alpha=1$, and an
exponential form when $\alpha=0$. When $\alpha$ is in the interval
$(0,1)$, the network exhibits a stretched exponential
distribution. We prove that the average topological distance
increases in a logarithmic scale of the network size, indicating
the existence of the small-world property. Furthermore, we obtain
the geographical edge-length distribution, the total geographical
length of all edges, and the average geographical distance of the
whole network. Interestingly, we found that the total edge-length
will sharply increase when $\alpha$ exceeds the critical value
$\alpha_c=1$, and the average geographical distance has an upper
bound independent of the network size. All the results are
obtained analytically with some reasonable approximations, which
are well verified by simulations.
\end{abstract}
\pacs{89.75.Hc, 87.23.Ge, 05.40.-a, 05.90.+m}

\maketitle

\section{Introduction}
Since the seminal works on the small-world phenomenon by Watts and
Strogatz \cite{Watts1998} and the scale-free property by
Barab\'asi and Albert \cite{Barabasi1999}, the studies of complex
networks have attracted a lot of interests within the physical
community
\cite{Albert2002,Dorogovtsev2002,Newman2003,Boccaletti2006}. Most
of the previous works focus on the topological properties (i.e.
non-geographical properties) of the networks. In this sense, every
edge is of length 1, and the \emph{topological distance} between
two nodes is simply defined as the number of edges along the
shortest path connecting them. To ignore the geographical effects
is reasonable for some networked systems (e.g. food webs
\cite{Garlaschelli2003}, citation networks \cite{Borner2004},
metabolic networks \cite{Jeong2000}), where the Euclidean
coordinates of nodes and the lengths of edges have no physical
meanings. Yet, many real-life networks, such as transportation
networks \cite{Sen2003,Guimera2005}, the Internet
\cite{Faloutsos1999,Pastor2001}, and power grids
\cite{Crucitti2004,Albert2004}, have well-defined node-positions
and edge-lengths. In addition to the topologically preferential
attachment introduced by Barab\'asi and Albert
\cite{Barabasi1999}, some recent works have demonstrated that the
spatially preferential attachment mechanism also plays a major
role in determining the network evolution
\cite{Yook2001,Barthelemy2003,MannaSen}.

Very recently, some authors have investigated the spatial
structures of the so-called \emph{optimal networks}
\cite{Gastner2006,Barthelemy2006}. An optimal network has a given
size and an optimal linking pattern, and is obtained by a certain
global optimization algorithm (e.g. simulated annealing) with an
objective function involving both geographical and topological
measures. Their works provide some guidelines in network design.
However, the majority of real networks are not fixed, but grow
continuously. Therefore, to study growing networks with an optimal
policy is not only of theoretical interest, but also of practical
significance. In this paper, we propose a growing network model,
in which, at each time step, one node is added and connected to
some existing nodes according to an optimal policy. The degree
distribution, edge-length distribution, and topological as well as
geographical distances are analytically obtained subject to some
reasonable approximations, which are well verified by simulations.

\section{Model}
Consider a square of size $1\times 1$ with open boundary
condition, that is, a open set $(0,1)\times (0,1)$ in Euclidean
space $\mathbb{R}^2$, where ``$\times$" signifies the Cartesian
product. This model starts with $m$ fully connected nodes inside
the square, all with randomly assigned coordinates. Since there
exists $m$ nodes initially, the discrete time steps in the
evolution are counted as $t=m+1, m+2, \cdots$. Then, at the $t$th
time step ($t>m$), a new node with randomly assigned coordinates
is added to the network. Rank each previously existing node
according to the following measure:
\begin{equation}
\omega_i=|\overrightarrow{r_i}-\overrightarrow{r_t}|^2/k_i^\alpha(t),\texttt{
}i=1,2,\cdots,t-1,
\end{equation}
and the node having the smallest $\omega$ is arranged on the top.
Here, each node is labelled by its entering time,
$\overrightarrow{r_t}$ represents the position of the $t$th node,
$k_i(t)$ is the degree of the $i$th node at time $t$, and $\alpha
\geq 0$ is a free parameter. The newly added node will connect to
$m$ existing nodes that have the smallest $\omega$ (i.e. on the
top of the queue). All the simulations and analyses shown in this
paper are restricted to the specific case $m=1$, since the
analytical approach is only valid for the tree structure with
$m=1$. However, we have checked that all the results will not
change qualitatively if $m$ is not too large compared with the
network size.

\begin{figure}
\scalebox{0.6}[0.6]{\includegraphics{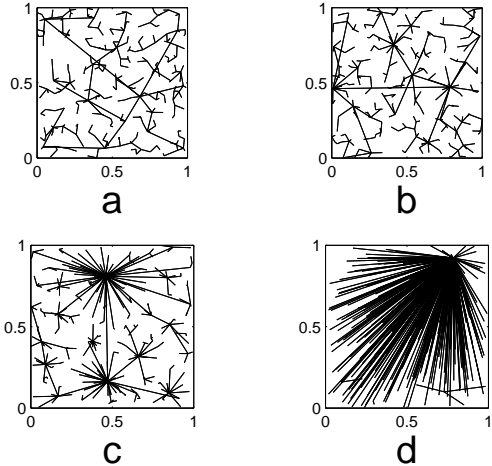}} \caption{Some
network examples for different values of $\alpha$: (a) $\alpha=0$,
(b) $\alpha=0.5$, (c) $\alpha=1$, (d) $\alpha=2$. All the four
networks are of size $N=300$ and $m=1$.}
\end{figure}

\begin{figure}
\scalebox{0.8}[0.8]{\includegraphics{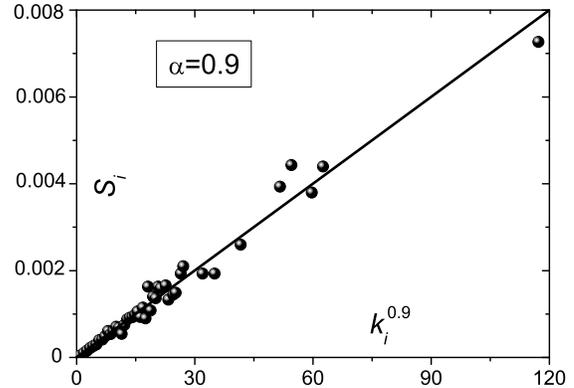}} \caption{Numerical
simulation about the relation between $S_i$ and $k_i$. The solid
line represents the assumption $S_i \propto k_i^\alpha$. In this
simulation, we first generate a network of size $N=10000$ by using
the present optimal growing policy. To detect $S_i$, some test
nodes with randomly assigned coordinates are used, and for each
test node, following Eq. (1), the existing nodes with minimal
$\omega$ is awarded one point. The test node will be removed from
the network after testing. The area $S_i$ for the $i$th node is
approximately estimated as the ratio of the score node $i$ has
eventually got after all the testings to the total number of test
nodes. The simulation shown here is obtained from 30000 test
nodes, and the parameter $\alpha=0.9$ is fixed.}
\end{figure}

In real geographical networks, short edges are always dominant
since constructing long edges will cost more \cite{Waxman1988}. On
the other hand, connecting to the high-degree nodes will make the
average topological distance from the new node to all the previous
nodes shorter. These two ingredients are described by the
numerator and denominator of Eq. (1), respectively. In addition,
the weights of these two ingredients are usually not equal. For
example, the airline travellers worry more about the number of
sequential connections \cite{Guimera2005}, the railway travellers
and car drivers consider more about geographical distances
\cite{Sen2003}, and the bus riders often simultaneously think of
both factors \cite{Zhang2006}. In the present model, if
$\alpha=0$, only the geographical ingredient is taken into
account. At another extreme, if $\alpha\rightarrow \infty$, the
geographical effect vanishes.

Fig. 1 shows some examples for different values of $\alpha$. When
only the geographical ingredient is considered ($\alpha=0$), most
edges are very short and the degree distribution is very narrow.
In the case of $\alpha=0.5$, the average geographical length of
edges becomes longer, and the degree distribution becomes broader.
When $\alpha=1$, the scale-free structure emerges and a few hub
nodes govern the whole network evolution. As $\alpha$ becomes very
large, the network becomes star-like.

\section{Degree Distribution}
At the $t$th time step, there are $t-1$ pre-existing nodes. The
square can be divided into $t-1$ regions such that if a new node
is fallen inside the $i$th region $S_i,\texttt{ }i\in
\{1,2,\cdots,t-1\}$, the quantity
$|\overrightarrow{r_i}-\overrightarrow{r_t}|^2/k_i^\alpha(t)$ is
minimized, thus the new node will attach an edge to the $i$th
node. Since the coordinate of the new node is randomly selected
inside the square, the probability of connecting with the $i$th
node is equal to the area of $S_i$. If the positions of nodes are
uniformly distributed, statistically, the area of $S_i$ is
approximately proportional to $k_i^\alpha(t)$ with a time factor
$h(t)$ as $S_i \sim h(t)k_i^\alpha(t)$. Fig. 2 shows the typical
simulation results, which strongly support the valid of this
assumption. Accordingly, by using the mean-field theory
\cite{Barabasi1999}, an analytic solution of degree distribution
can be obtained. However, when $\alpha>1$, most edges are
connected to one single node (see Fig. 1d), so analytic solution
is unavailable. Here, we only consider the case of $1\geq \alpha
\geq 0$. Assume
\begin{equation}
\sum_{i=1}^t k_i^\alpha(t)\approx At,
\end{equation}
where $A$ is a constant that can be determined self-consistently.
Using the continuum approximation in time variable $t$, the
evolving of node $i$'s degree reads
\begin{equation}
{dk_i(t)\over dt}=\frac{S_i}{\sum^t_iS_i}={k_i^\alpha\over At},
\end{equation}
with the initial condition $k_i(i)=1$. The solution is
\begin{equation}
k_i(t)=F(t/i),
\end{equation}
where
\begin{equation}
F(x)=\left(1+{1-\alpha\over A}\ln(x) \right) ^{1/(1-\alpha)}.
\end{equation}
Accordingly, the degree distribution can be obtained as
\begin{equation}
P(k)=\frac{1}{t}\sum^t_{i=1}\delta(k_i(t)-k)={A\over k^\alpha
e^{{A(k^{1-\alpha}-1)\over 1-\alpha}}}.
\end{equation}
The constant $A=\int_0^1 du F^{\alpha}(1/u)$ is determined by the
condition $\int kP(k) dk= \langle k\rangle =2$, where $\langle
k\rangle=2m=2$ signifies the average degree. The above solution is
similar to the one obtained by using the approach of \emph{rate
equation} proposed by Krapivsky \emph{et al.}
\cite{krapivsky2000}. In addition, one should note that if
$\alpha=1$, the mean-field theory yields a solution $P(k) \sim
k^{-3}$, which is comparable to the exact analytic solution
$P(k)=4/(k(k+1)(k+2))$.

Clearly, the degree distribution obeys a power-law form at
$\alpha=1$, and an exponential form at $\alpha=0$. When $\alpha$
is in the interval $(0,1)$, the networks display the so-called
stretched exponential distribution \cite{Laherrere1998}: For small
$\alpha$, the distribution is close to an exponential one, while
for large $\alpha$, it is close to a power law. This result is in
accordance with the situation of transportation networks. If only
the geographical ingredient is taken into account (e.g. road
networks \cite{Gastner2006}), then the degree distribution is very
narrow. On the contrary, if the topological ingredient plays a
major role (e.g. airport networks \cite{Guimera2005}), then the
scale-free property emerges. When both the two ingredients are not
neglectable (e.g. bus networks \cite{Zhang2006}), the degree
distribution is intervenient between power-law and exponential
ones.

\begin{figure}
\scalebox{0.8}[0.8]{\includegraphics{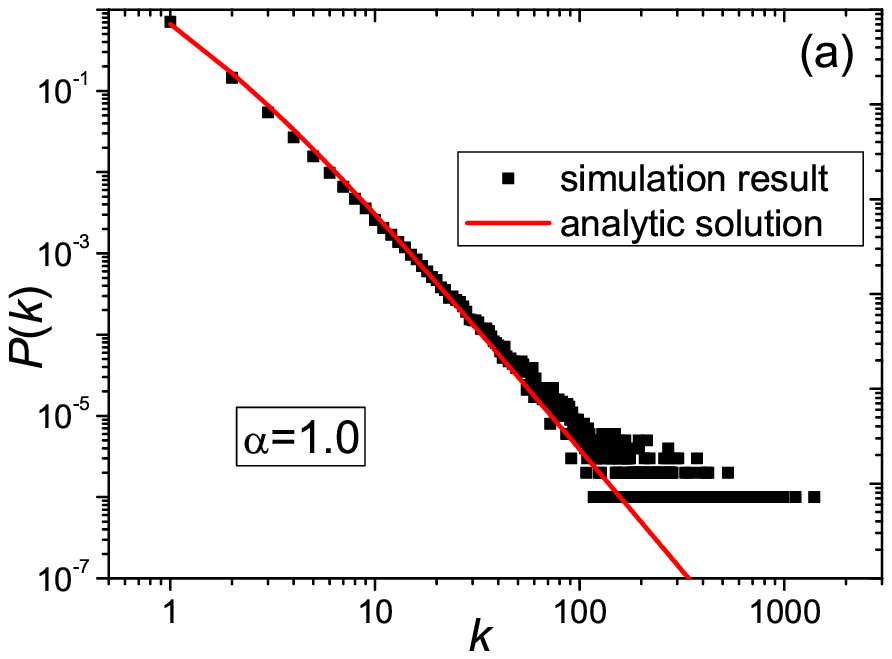}}
\scalebox{0.8}[0.8]{\includegraphics{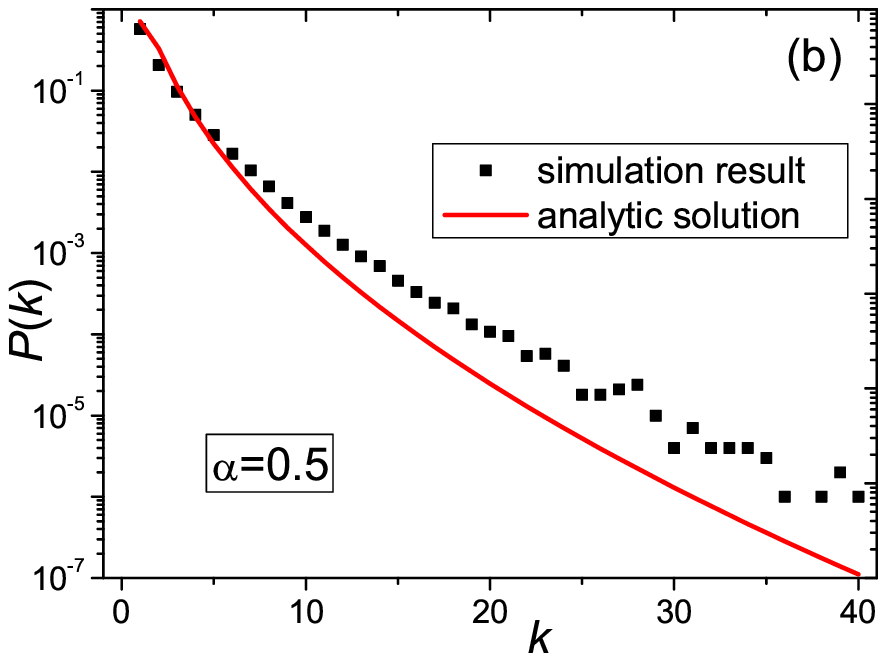}} \caption{(Color
online) The degree distributions for the cases of $\alpha=1$ (a)
and $\alpha=0.5$ (b). The black squares and red curves represent
simulated and analytic results, respectively. All the data are
averaged over $100$ independent runs, with network size $N=10000$
(i.e. $t=10000$) fixed.}
\end{figure}

Fig. 3 shows the simulation results for $\alpha=1$ and
$\alpha=0.5$. The degree distribution follows a power-law form
when $\alpha=1$, which well agrees with the analytic solution. In
the case of $\alpha=0.5$, the degree distribution is more
exponential. However, it is remarkably broader than that of the
Erd\"os-R\'enyi model \cite{Erdos1960}. Note that, the positions
of all the nodes are not completely uniformly distributed, which
will affect the degree distribution. This effect becomes more
prominent when the geographical ingredient plays a more important
role (i.e. smaller $\alpha$). Therefore, although the simulation
result for $\alpha=0.5$ is in accordance with the analysis
qualitatively, the quantitative deviation can be clearly observed.

\section{Topological Distance}
Denote by $l(i)$ the topological distance between the $i$th node
and the first node. By using mathematical induction, we can prove
that there exists a positive constant $M$, such that
$l(t)<M\texttt{ln}(t)$. This proposition can be easily transferred
to prove the inequality $l(t+1)<M\texttt{ln}(t+1)$ under the
condition $l(i)<M\texttt{ln}(i)$ for $1\leq i\leq t$. Indeed,
since the network has a tree structure, $l(i)$ does not depend on
time $t$. Under the framework of the mean field theory, the
iteration equation for $l(t)$ reads
\begin{equation}
l(t+1)={\sum_{1\leq i\leq t} k^\alpha_i(t) l(i)\over A t}+1,
\end{equation}
with the initial condition $l(1)=0$. Eq. (7) can be understood as
follows: At the $(t+1)$th time step, the $(t+1)$th node has
probability $k_i^\alpha(t)/A t$ to connect with the $i$th node.
Since the average topological distance between the $i$th node and
the first node is $l(i)$, the topological distance of the
$(t+1)$th node to the first one is $l(i)+1$ if it is connected
with the $i$th node. According to the induction assumption,
\begin{equation}
l(t+1)<{M\sum_{1\leq i\leq t} k^\alpha_i(t) \texttt{ln}(i)\over
At}+1.
\end{equation}
Note that, statistically, $k_i(t)>k_j(t)$ if $i<j$, therefore
\begin{equation}
\sum_{1\leq i\leq t}k^\alpha_i(t) \texttt{ln}(i)<\sum_{1\leq i\leq
t}\langle k^\alpha(t)\rangle\texttt{ln}(i)=\sum_{1\leq i\leq
t}A\texttt{ln}(i),
\end{equation}
where $\langle \cdot \rangle$ denotes the average over all the
nodes. Substituting inequality (9) into (8), we have
\begin{equation}
l(t+1)<\frac{M}{t}\sum_{1\leq i\leq t}\texttt{ln}(i)+1.
\end{equation}
Rewriting the sum in continuous form, we obtain
\begin{equation}
l(t+1)<\frac{M}{t}\int_1^t\texttt{ln}(x)\texttt{d}x+1<M\texttt{ln}(t+1).
\end{equation}
According to the mathematical induction principle, we have proved
that the topological distance between the $i$th node and the first
node, denoted by $l(i)$, could not exceed the order
$O(\texttt{ln}(i))$. For arbitrary nodes $i$ and $j$, clearly, the
topological distance between them could not exceed the sum
$l(i)+l(j)$, thus the average topological distance $\langle d
\rangle$ of the whole network could not exceed the order
$O(\texttt{ln}(i))$ either. This topological characteristic is
referred to as the small-world effect in network science
\cite{Watts1998}, and has been observed in a majority of real
networks. Actually, one is able to prove that the order of $l(t)$
in the large $t$ limit is equal to $\texttt{ln}(t)$ (see Appendix
A for details).

Furthermore, the iteration equation
\begin{equation}
l(t+1)={\sum_{i=1}^t f(i)l(i)\over\sum_{i=1}^t f(i)}+a(t),
\end{equation}
for general functions $f(i)$ and $a(i)$, has the following
solution:
\begin{equation}
l(t+1)=l(1)+a(t)+\sum_{j=1}^{t-1}{f(j+1)a(j)\over \sum_{i=1}^{j+1}
f(i)}.
\end{equation}
For the two special cases of $\alpha=0$ ($a(i)=1$, $f(i)=1$) and
$\alpha=1$ ($a(i)=1$, $f(i)=1/\sqrt{i}$), the solutions are simply
$l(t)=\texttt{ln}(t)$ and $l(t)=\frac{1}{2}\texttt{ln}(t)$,
respectively.

In Fig. 4, we report the simulation results about the average
distance $\langle d \rangle$ vs network size $t$. In each case,
the data points can be well fitted by a straight line in the
semi-log plot, indicating the growth tendency $\langle d \rangle
\sim \texttt{ln}(t)$, which agrees well with the analytical
solution.

\begin{figure}
\scalebox{0.8}[0.8]{\includegraphics{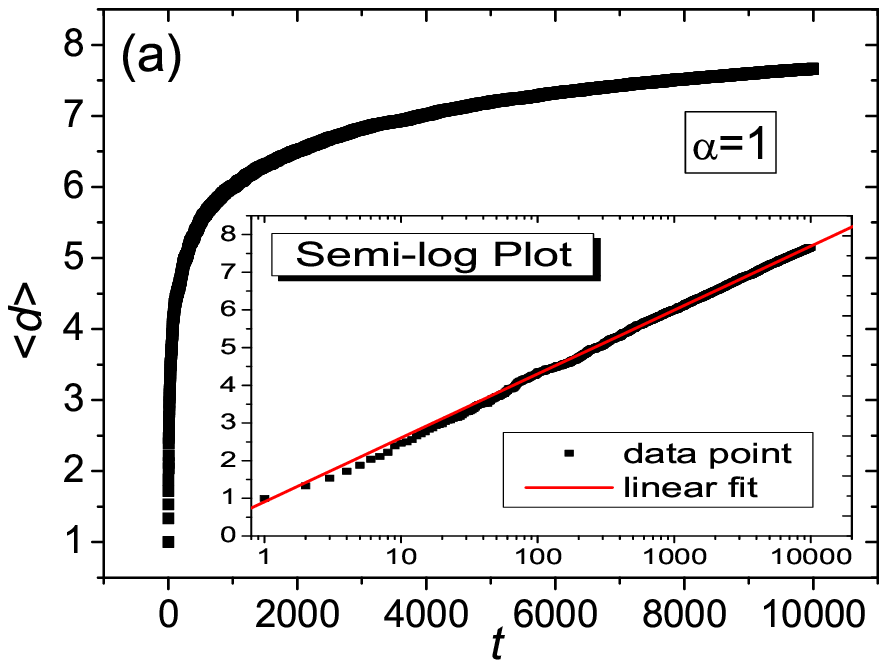}}
\scalebox{0.8}[0.8]{\includegraphics{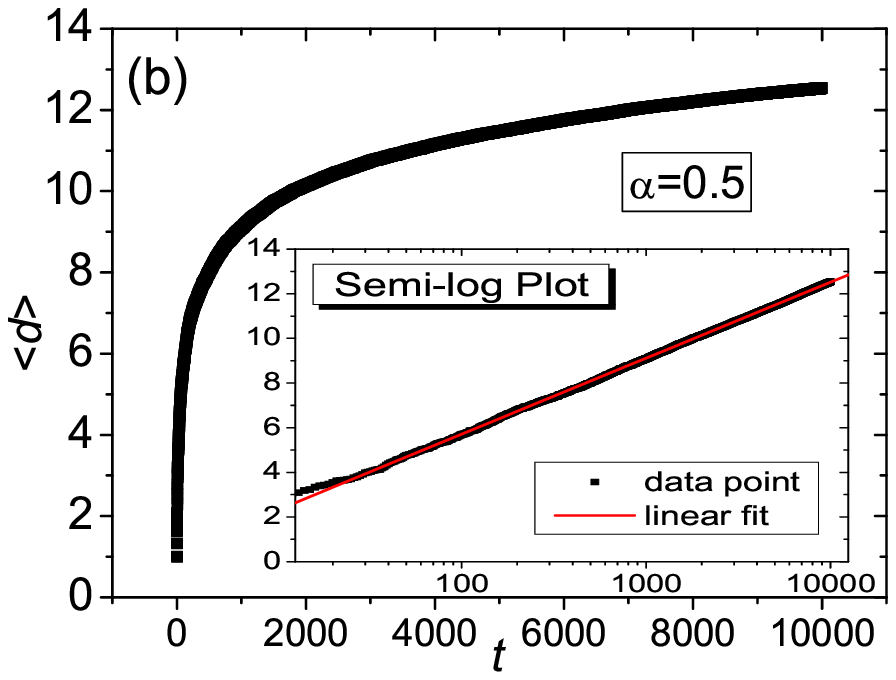}} \caption{(Color
online) The average distance vs network size for the cases of
$\alpha=1$ (a) and $\alpha=0.5$ (b). As shown in each inset, the
data points can be well fitted by a straight line in the semi-log
plot, indicating the growth of the average distance approximately
obeys the form $\langle d \rangle \sim \texttt{ln}(t)$. All the
data are averaged over $5$ independent runs, where the maximal
network size is $N=10000$ (i.e. $t=10000$).}
\end{figure}

\section{Edge Length Distribution}
Denote by $e_{ij}$ the edge between nodes $i$ and $j$, and the
geographical length of edge $e_{ij}$ is
$r_{ij}=|\overrightarrow{r_i}-\overrightarrow{r_j}|$. When the
$(N+1)$th node is added to the network, the geographical length of
its attached edge approximately obeys the distribution
\begin{equation}
Q(r)=2\pi r \sum_{i=1}^N e^{-\pi{r^2\over k_i^\alpha}\sum_j
k_j^\alpha},
\end{equation}
where $r\ll 1$ in the large $N$ limit. The derivation of this
formula is described as follows. The probability of the edge
length being between $r$ and $r+dr$ is given by the summation
$\sum_iQ_idr$, where $Q_idr$ is the probability that $r_{i,N+1}$
falls between $r$ and $r+dr$, and the node $i$ minimizes the
quantity
$|\overrightarrow{r_{N+1}}-\overrightarrow{r_i}|^2/k_i^\alpha$
among all the $N$ previously existing nodes. This probability is
approximately given by
\begin{eqnarray}
Q_i&\approx& 2\pi r \prod_j\left(1- {\pi r^2 k_j^\alpha\over
k_i^\alpha}\right)
\nonumber\\
&\approx& 2\pi r e^{-\pi{r^2\over k_i^\alpha}\sum_j k_j^\alpha}.\nonumber\\
\end{eqnarray}
Straightforwardly, the geographical length distribution of the
newly added edge at the $t$th time step (the $t$th edge for short)
is obtained as
\begin{eqnarray}
P(t,r)&=& 2\pi r \sum_i e^{-\pi{r^2\over k^\alpha_i(t)}\sum_j k^\alpha_j(t)}
\nonumber\\
&\approx& 2\pi r t\int_{1/t}^1 du e^{-\pi At r^2/F^\alpha(1/u)}
\nonumber\\
&\approx& 2\pi r t \int_0^1 du e^{-\pi At r^2/F^\alpha(1/u)}.
\end{eqnarray}
The lower boundary in the integral is replaced by 0 in the last
step, which is valid only when $\alpha <1$. The cumulative length
distribution of the edges at time step $T$ is given by
\begin{eqnarray}
P(r) &=& \frac{1}{T}\int_1^T P(t,r)dt
\nonumber\\
&=& {2\over \pi r^3 A^2T}\int_0^1 du F^{2\alpha} \bigg[e^{-\pi
Ar^2/F^\alpha}\left(1+{\pi Ar^2\over F^\alpha}\right)
\nonumber\\
&& -e^{-\pi ATr^2/F^\alpha}\left(1+{\pi ATr^2\over
F^\alpha}\right) \bigg],
\end{eqnarray}
where the argument of function $F$ is $1/u$. For $1/\sqrt{T}\ll
r\ll 1$, the approximate formula for $P(r)$ reads
\begin{equation}
P(r)\approx {2\over\pi r^3 A^2T}\int_0^1 du
F^{2\alpha}\left({1\over u}\right)
\end{equation}
and, when $r\ll 1/\sqrt{T}$,
\begin{equation}
P(r)\approx \pi rT.
\end{equation}
If $\alpha=1$, the last step in Eq. (16) is invalid but the
analytic form for $P(t,r)$ can be directly obtained as
\begin{eqnarray}
P(t,r)=2\pi r t\int_{1/t}^1 du e^{-2\pi tr^2\sqrt{u}}
\nonumber\\={1\over \pi r^3 t}[(1+2\pi r^2\sqrt{t})e^{-2\pi
r^2\sqrt{t}} -(1+2\pi r^2 t)e^{-2\pi r^2 t}]
\end{eqnarray}
Therefore, when $1/T^{1/4}\ll r\ll 1$, $P(r)$ is approximately
given by
\begin{equation}
P(r)\approx {1\over \pi r^3T}\ln{C\over 2\pi r^2},
\end{equation}
where $C$ is a numerical constant, and when $r\ll 1/\sqrt{T}$,
$P(r)$ has the same form as that of Eq. (19).

Fig. 5 plots the cumulative edge-length distributions. From this
figure, one can see a good agreement between the theoretical and
the numerical results. Furthermore, one can calculate the expected
value of the $t$th edge's geographical length, as
\begin{equation}
\overline{r}(t)={\int_0^1 rP(t,r)dr\over\int_0^1 P(t,r) dr}
={1\over 2\sqrt{At}}{\int_0^1 du F^{3\alpha/2}(1/u) \over\int_0^1
du F^\alpha(1/u)},
\end{equation}
which is valid only for sufficiently large $t$ and $\alpha<1$.
According to Eq. (22), $\overline{r}(t)$ decreases as $1/\sqrt{t}$
as $t$ increases, which is consistent with the intuition since all
the $t$ nodes are embedded into a 2-dimensional Euclidean space.
It may also be interesting to calculate the total length $R(t)$ of
all the edges at the time step $t$, as
\begin{equation}
R(t)=\sum^t_{i=1}\overline{r}(i)\approx \int_1^t dt'
\overline{r}(t')=\sqrt{{t\over A}} {\int_0^1 du F^{3\alpha/2}(1/u)
\over\int_0^1 du F^\alpha(1/u)}.
\end{equation}
$R(t)$ is proportional to $\sqrt{t}$ for $1>\alpha>0$. When
$\alpha>1$, a finite fraction of nodes will be connected with a
single hub node and therefore we expect that $R(t)\sim t$ in this
case. Therefore, in the large $t$ limit, $R(t)$ will increase
quite abruptly when the parameter $\alpha$ exceeds 1. This
tendency is indeed observed in our numerical simulations, as shown
in Fig. 6.

\begin{figure}
\scalebox{0.8}[0.8]{\includegraphics{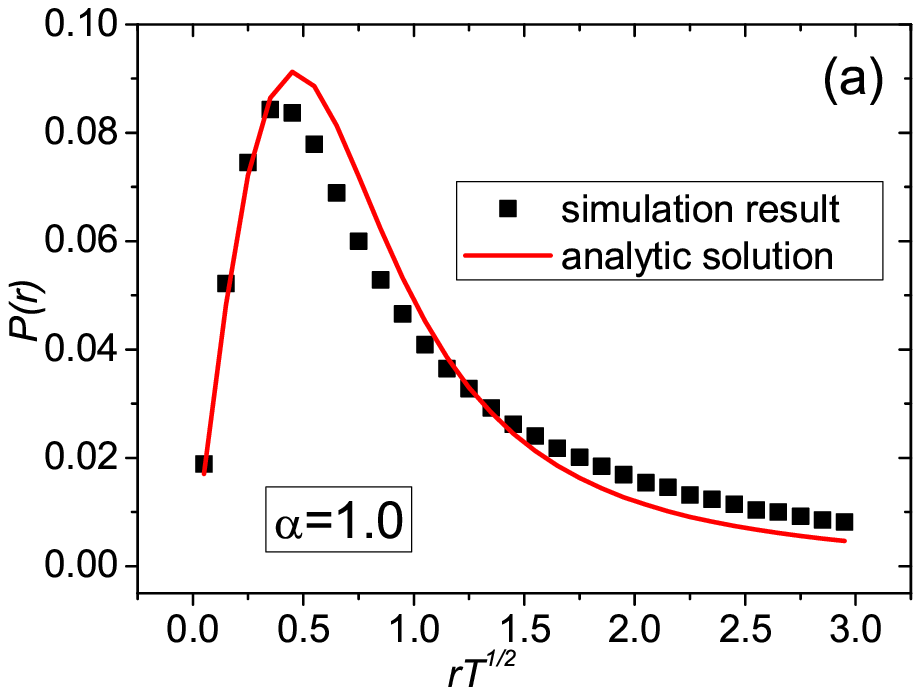}}
\scalebox{0.8}[0.8]{\includegraphics{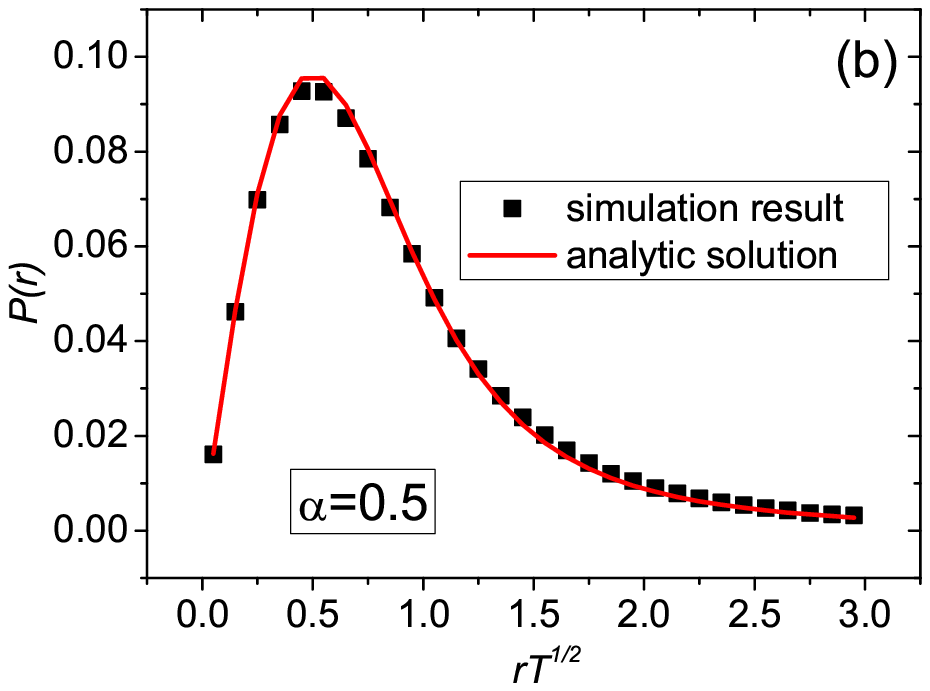}} \caption{(Color
online) The edge length distributions for the cases of $\alpha=1$
(a) and $\alpha=0.5$ (b). The black squares and red curves
represent simulation and analytic results, respectively. All the
data are averaged over $100$ independent runs, with network size
$N=10000$ (i.e. $T=10000$) fixed.}
\end{figure}

\section{Geographical Distance}
For an arbitrary path $i_0\rightarrow i_1\rightarrow \cdots
\rightarrow i_n$ from node $i_0$ to $i_n$, the corresponding
geographical length is $\sum_{u=0}^{n-1}r_{i_ui_{u+1}}$, where
$r_{ij}$ denotes the length of edge $e_{ij}$. Accordingly, the
geographical distance between two nodes is defined as the minimal
geographical length of all the paths connecting them. Now, we
calculate the geographical distance $s(i)$ between the $i$th node
and the first node. Since our network is a tree graph, $s(i)$ does
not depend on time. By using the mean field theory, we have
\begin{equation}
s(t+1)={\sum_{i\le t} k^\alpha_i(t) s(i)\over A
t}+\overline{r}(t+1),
\end{equation}
or
\begin{equation}
s(t+1)={1\over A}\int_{1/t}^1 du F^\alpha\left({1\over
u}\right)s(ut)+{W\over\sqrt{t+1}},
\end{equation}
where, according to Eq. (22),
\begin{equation}
W={1\over 2\sqrt{A}}{\int_0^1 du F^{3\alpha/2}(1/u) \over\int_0^1
du F^\alpha(1/u)}.
\end{equation}

It is not difficult to see that $s(t)$ has an upper bound as $t$
approaches infinity. One can use the trial solution
$s(t)=B-C/t^\beta$ to test this conclusion:
\begin{equation}
B-C/(t+1)^\beta=B-CE/(At^\beta)+W/\sqrt{t+1},
\end{equation}
where
\begin{equation}
E=\int_{1/t}^1 du F^\alpha(1/u)(1/u^\beta).
\end{equation}
From Eq. (27), one obtains that $\beta=1/2$.

Similar to the solution of Eq. (13), $s(t)\to 1$ as $t\to\infty$
for $\alpha=0$ and $s(t)\to 1/\sqrt{2}$ as $t\to\infty$ for
$\alpha=1$. However, it only reveals some qualitative property,
and the exact numbers are not meaningful. This is because the
value of $s(t)$ is obtained by the average over infinite
configurations for infinite $t$, while in one evolving process
$s(t)$ is mainly determined by the randomly assigned coordinates
of the $t$th node.

\begin{figure}
\scalebox{0.8}[0.8]{\includegraphics{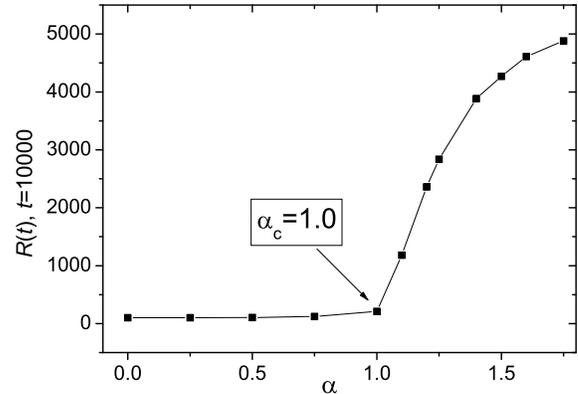}} \caption{The total
edge length $R(t)$ vs parameter $\alpha$ with $t=10000$ fixed.
$R(t)$ sharply increases when $\alpha$ exceeds the critical value
$\alpha_c=1$, which well agrees with the theoretical prediction.}
\end{figure}

\section{Conclusion and Discussion}
In many real-life transportation networks, the geographical effect
can not be ignored. Some scientists proposed certain global
optimal algorithms to account for the geographical effect on the
structure of a static network \cite{Gastner2006,Barthelemy2006}.
On the other hand, many real networks grow continuously.
Therefore, we proposed a growing network model based on an optimal
policy involving both topological and geographical measures. We
found that the degree distribution will be broader when the
topological ingredient plays a more important role (i.e. larger
$\alpha$), and when $\alpha$ exceeds a critical value
$\alpha_c=1$, a finite fraction of nodes will be connected with a
single hub node and the geographical effect will become
insignificant. This critical point can also be observed when
detecting the total geographical edge-length $R(t)$ in the large
$t$ limit. We obtained some analytical solutions for degree
distribution, edge-length distribution, and topological as well as
geographical distances, based on reasonable approximations, which
are well verified by simulations.

\begin{table}
\caption{Empirical degree distributions for geographical
transportation networks.}
\begin{ruledtabular}
\begin{tabular}{lcr}
Data-Class & Data-Set & Degree Distribution\\
\hline
Power Grid & Southern California \cite{Amaral2000} & Exponential\footnote{In Ref. \cite{Barabasi1999}, the authors claimed that this distribution follows a power-law form $p(k)\sim k^{-\gamma}$ with exponent $\gamma \approx 4$. Actually, when $\gamma$ gets larger, the network will topologically gets closer to random graph \cite{Braunstein2003}. }\\
  & Whole US \cite{Albert2004} & $p(k)\sim e^{-0.5k}$\\
  & Italian \cite{Crucitti2004} & $p(k)\sim e^{-0.55k}$\\
\hline
Subway & Boston \cite{Latora2002} & Narrow: $k_{\texttt{max}}<10$\\
       & Seoul \cite{Chang2006} & Narrow: $k_{\texttt{max}}<10$\\
       & Tokyo \cite{Chang2006} & Narrow: $k_{\texttt{max}}<10$\\
\hline
Railway & Indian \cite{Sen2003} & $p(k)\sim e^{-0.085k}$\\
        & Switzerland \cite{Kurant2006} & Exponential\\
        & Central Europe \cite{Kurant2006} & Exponential\\
\hline
Bus \& Tramway & Krak\'ow \cite{Sienkiewicz2005} & Narrow: $k_{\texttt{max}}=11$\\
               & Warsaw \cite{Sienkiewicz2005} & Narrow: $k_{\texttt{max}}=13$\footnote{In Ref. \cite{Kurant2006}, the authors displayed an exponential degree distribution of mass transportation networks in Warsaw.}\\
               & Szczecin \cite{Sienkiewicz2005} & Narrow: $k_{\texttt{max}}=18$\\
               & Bia\l ystok \cite{Sienkiewicz2005} & Narrow: $k_{\texttt{max}}=19$\\
\hline
Airport & World-Wide \cite{Guimera2005} & $p(k)\sim k^{-2.0}$\footnote{Actually a truncated power-law distribution.}\\
        & Chinese \cite{Liu2006} & $p(k)\sim k^{-2.05}$\footnote{Actually a double power-low distribution.}\\
        & Indian \cite{Bagler2004} & $p(k)\sim k^{-2.2}$\\

\end{tabular}
\end{ruledtabular}
\end{table}

Although the present model is based on some ideal assumptions, it
can, at least qualitatively, reproduce some key properties of the
real transportation networks. In Table 1, we list some empirical
degree distributions of transportation networks. Clearly, when
building a new airport, we tend to firstly open some flights
connected with previously central airports which are often of very
large degrees. Even though the central airports may be far from
the new one, to open a direct flight is relatively convenient
since one doesn't need to build a physical link. Therefore, the
geographical effect is very small in the architecture of airport
networks, which corresponds to the case of larger $\alpha$ that
leads to an approximately power-law degree distribution. For other
four cases shown in Table 1, a physical link, which costs much, is
necessary if one wants to connect two nodes, thus the geographical
effect plays a more important role, which corresponds to the case
of smaller $\alpha$ that leads to a relatively narrow
distribution.

A specific measure of geographical network is its edge-length
distribution. A very recent empirical study \cite{Gastner2006}
shows that the edge-length distribution of the highly heterogenous
networks (e.g. airport networks, corresponding to the present
model with larger $\alpha$) displays a single-peak function with
the maximal edge-length about five times longer than the peaked
value (see Fig. 1c of Ref. \cite{Gastner2006}), while in the
extreme homogenous networks (e.g. railway networks, corresponding
to the present model with $\alpha\rightarrow 0$), only the very
short edge can exist (see Fig. 1a of Ref. \cite{Gastner2006}).
These empirical results agree well with the theoretical
predictions of the present model. Firstly, when $\alpha$ is
obviously larger than zero, the edge-length distribution is
single-peaked with its maximal edge-length about six times longer
than the peak value (see Fig. 5). And, when $\alpha$ is close to
zero, Eq. (17) degenerates to the form
\begin{equation}
P(r)=\frac{2}{\pi r^3A^2N},
\end{equation}
where $N$ denotes the network size. Clearly, in the large $N$
limit, except a very few initially generated edges, only the edge
of very small length $r$ can exist.

The analytical approach is only valid for the tree structure with
$m=1$. However, we have checked that all the results will not
change qualitatively if $m$ is not too large compared with the
network size. Some analytical methods proposed here are simple but
useful, and may be applied to some other related problems about
the statistical properties of complex networks. For example, a
similar (but much simpler) approach, taken in section 4, can also
be used to estimate the average topological distance for some
other geographical networks \cite{Zhou2005,ZhangZZ2006}.

Finally, it is worthwhile to emphasize that, the geographical
effects should also be taken into account when investigating the
efficiency (e.g. the traffic throughput \cite{Yan2006}) of
transportation networks. Very recently, some authors started to
consider the geographical effects on dynamical processes, such as
epidemic spreading \cite{Xu2006} and cascading \cite{Huang2006},
over scale-free networks. We hope the present work can further
enlighten the readers on this interesting subject.

\begin{acknowledgments}
The authors wish to thank Dr. Hong-Kun Liu to provide us some very
helpful data on Chinese city-airport networks. This work was
partially supported by the National Natural Science Foundation of
China under Grant Nos. 10635040, 70471033, and 10472116, the
Special Research Founds for Theoretical Physics Frontier Problems
under Grant No. A0524701, and Specialized Program under the
Presidential Funds of the Chinese Academy of Science.
\end{acknowledgments}

\appendix

\section{The solution of $l(t)$}

Substituting Eq. (4) into Eq. (7), one obtains that
\begin{equation}
l(t+1)={1\over A}\int_{1/t}^1 du F^\alpha\left({1\over
u}\right)l(ut)+1.
\end{equation}
Then, define
\begin{equation}
B=A\left\{\int_0^1 du F^\alpha(1/u)\ln (1/u)\right\}^{-1}.
\end{equation}
We next prove that $l(t)\approx B\ln t$ in the large $t$ limit by
using mathematical induction. Suppose for sufficiently large $t$,
all $l(i)$ are less than $C\ln (i)$ for $i\leq t$ with $C$ being a
constant greater than $B$. Then, from Eq. (A1), we have
\begin{eqnarray}
l(t+1)&\leq& {C\over A}\int_{1/t}^1 du F^\alpha\left({1\over
u}\right) \ln
(ut)+1 \nonumber\\
&=& -{C\over B}+1+M\ln (t)+O\left({1\over t}\right)\nonumber\\
&\leq& C\ln (t+1)
\end{eqnarray}
Therefore, $l(i)\leq C\ln (i)$ for all $i$.  Similarly, suppose
for sufficiently large $t$, all $l(i)$ are greater than $Q\ln i$
for $i\leq t$ with $Q$ being a constant less than $B$. Then, from
Eq. (A1), we have
\begin{eqnarray}
l(t+1)&\geq& {Q\over A}\int_{1/t}^1 du F^\alpha \left({1\over
u}\right) \ln
(ut)+1 \nonumber\\
&=& -{Q\over B}+1+M\ln (t)+O\left({1\over t}\right)\nonumber\\
&\geq& Q\ln (t+1).
\end{eqnarray}
Therefore, $l(i)\geq Q\ln (i)$ for all $i$.

Combine both the upper bound (A3) and lower bound (A4), we obtain
the order of $l(t)$ in the large $t$ limit, as $l(t)\approx B\ln
(t)$.

\end{document}